\documentclass{PoS}
\usepackage[authoryear]{natbib}
\bibpunct{(}{)}{;}{a}{}{,}
\newcommand {\ga} {\ {\raise-.5ex\hbox{$\buildrel>\over\sim$}}\ }
\newcommand {\la} {\ {\raise-.5ex\hbox{$\buildrel<\over\sim$}}\ }

\title{Kinematics and Dynamics of kiloparsec-scale Jets in Radio
    Galaxies with SKA}

\ShortTitle{Large-scale Extragalactic Jets}

\author{\speaker{R.A. Laing}
        \\
        ESO, Karl-Schwarzschild-Stra\ss e 2, D-85748 Garching-bei-M\"unchen, Germany\\
        E-mail: \email{rlaing@eso.org}}

\abstract{We explore the use of SKA to deduce the physical parameters
  of kiloparsec-scale jet flows in radio galaxies. Jets in Active Galactic Nuclei are
  relativistic where they are first formed, but their speeds and
  compositions change as they propagate.  It has long been known that
  kiloparsec-scale jets in radio galaxies can be divided into two
  flavours: strong (found in powerful sources, narrow and terminating
  in compact hot-spots) and weak (found in low-luminosity sources,
  geometrically flaring, unable to form hot-spots and terminating in diffuse lobes
  or tails).  We have developed methods to model AGN jets as
  intrinsically symmetrical, relativistic flows by fitting to deep,
  well-resolved radio images in Stokes $I$, $Q$ and $U$.  This has
  yielded a wealth of information about the brightest few weak-flavour
  jets. Our first key objective is to observe large samples of weak
  and transition jets at 0.1 -- 0.5 arcsec resolution with SKA1-MID. This would
  allow us to see how jet propagation depends on power and environment
  and to quantify the energy and momentum input into the IGM.  We will
  require typical noise levels of 1$\mu$Jy/beam, and may be able to
  exploit survey imaging in some cases.  Our second, more challenging,
  application is to determine the velocity fields in strong-flavour
  jets. Do they have very fast ($\Gamma \approx 5 - 10$) spines?  Is
  there evidence for magnetic confinement by a toroidal field?  What
  are their energy fluxes? This is a major imaging challenge for SKA2:
  we need resolution better than 0.05 arcsec, ideally in the 1 --
  10\,GHz frequency range, with rms noise levels $\approx$10\,nJy/beam
  and extremely high dynamic range, imaging fidelity and polarization
  purity.}

\FullConference{Advancing Astrophysics with the Square Kilometre Array,\\
		June 9-13, 2014\\
		Giardini Naxos, Sicily, Italy }

\begin{document}

\section{Introduction}
\label{whybother}

Jets from active galactic nuclei (AGN) are important in many areas of
astrophysics: they extract energy from supermassive black holes,
produce the most energetic photons (and perhaps cosmic rays) we
observe, act as conveyors of ultrarelativistic particles and magnetic
fields from the parsec-scale environments of AGN to the
multi-kiloparsec scales of extended radio galaxies and quasars, and
supply copious amounts of energy to their surroundings, thereby
preventing cooling and profoundly affecting the evolution of massive
galaxies and clusters.

An  overview  of  the  study  of   AGN  jets  with  SKA  is  given  by
\cite{Agudo}. This paper  covers one aspect -- the physics  of jets on
kpc  scales --  in more  detail, emphasising  a number  of key  topics
(cf.\ \citealt{Blandford08}), as follows.
\begin{enumerate}
\item What are the velocity fields? Why do some jets expand rapidly,
  increase in brightness and decelerate while others remain fast and
  collimated? How does this relate to the origin of the Fanaroff-Riley
  classes \citep{FR74}?
\item How does jet composition change with distance from the AGN? What
  are the typical mass fluxes and entrainment rates?
\item What is the magnetic-field topology on kpc scales -- ordered or
  disordered?
\item How is jet propagation determined by the external environment
  (density, pressure, temperature, magnetization)?
\item How are jets confined: by external gas pressure, magnetic
  fields, some other mechanism or not at all?
\item Where and by what process are relativistic particles accelerated?
\item What are the jet energy and momentum fluxes?
\item What effects do the jets have on their surrounding IGM/ICM? Do
  they lead to shocks and heating? What effect do they have on
  external magnetic fields and hence on thermal transport?
\end{enumerate}
It has long been known that AGN radio jets divide into strong and weak
`flavours' \citep{Bridle}.  Strong-flavour jets are usually one-sided
($>$4:1) and well-collimated over their whole lengths and often
terminate in compact hot spots.  Weak-flavour jets are initially
asymmetric, but tend to symmetrize far from the nucleus. They also
`flare' with increasing distance from the AGN in two distinct
senses: {\it geometrically} (a significant increase in apparent
spreading rate) and {\it in brightness} (an increase in apparent
intensity, often following an initial `gap', or extended region in
which the radio emission is weak or undetectable).  The qualitative
picture is that both jet flavours are initially relativistic, but
weak-flavour jets decelerate to subrelativistic speeds on kpc scales,
while strong-flavour jets remain highly relativistic until they reach
the outer parts of the sources.
 
Thus far, we have derived quantitative models only for a small number
of {\em weak-flavour jets} using deep VLA observations.  {\em
  Strong-flavour jets} are harder to study as they are narrower than
weak-flavour jets (thus harder to resolve) and their counter-jets are
faint and difficult to isolate from surrounding lobe emission.  After
outlining the modelling methods and associated imaging requirements
(Section~\ref{models}), we therefore consider in turn the use of SKA1-MID 
to observe large samples of weak-flavour jets (Section~\ref{weak}) and
SKA2 to model strong-flavour jets for the first time with our methods 
(Section~\ref{powerful}).

\section{Measuring the flow parameters of jets}
\label{models}

Although numerical (GRMHD) simulations of jet formation have made
considerable progress recently (e.g.\ \citealt{mckinney}), the problem of
simulating the propagation of a very light, relativistic, magnetized
jet in three dimensions is computationally prohibitive, with poorly
known initial conditions: no simulation can yet hope to follow a jet
all the way from its formation on scales comparable with the
gravitational radius of the central black hole to the kiloparsec
scales which SKA can best resolve.  A more empirical approach to
estimation of jet flow parameters is therefore needed.

One powerful method is based on the realization that kpc-scale jets
are both significantly relativistic and intrinsically symmetrical, in
the specific sense that {\em apparent asymmetries due to relativistic
  aberration are much larger than any intrinsic asymmetries}. Jets
emit (often highly) polarized synchrotron radiation.  With the
assumptions that the flows are (on average) axisymmetric,
intrinsically symmetrical and stationary, the observed brightness
distributions in Stokes $I$, $Q$ and $U$ can be modelled to fit for
geometry, velocity field, intrinsic emissivity variation and
three-dimensional magnetic field configuration.  The method was first
described in \cite{LB02a}; a comprehensive overview is given by
\cite{LB14}.  The key to the method is to determine the velocity and
inclination angle independently by comparing emission from the main
and counter-jets in {\em both total intensity and linear
  polarization}.  The observed polarizations in the approaching and
receding jets are not the same because they are effectively observed at different
angles to the line of sight in the rest frame, $\theta^\prime$. For
antiparallel jets at angle $\theta$ to the line of sight in the
observed frame,
\begin{eqnarray*}
\sin\theta^\prime_{\rm j} &=& [\Gamma(1-\beta\cos\theta)]^{-1}\sin\theta \\
\sin\theta^\prime_{\rm cj} &=& [\Gamma(1+\beta\cos\theta)]^{-1}\sin\theta \\
\end{eqnarray*} 
where subscripts j and cj refer to the approaching and receding jets,
$\beta$ is the flow velocity in units of $c$ and $\Gamma$ is the bulk
Lorentz factor.  $\beta$ and $\theta$ cannot be determined
independently from images of total intensity alone: the use of linear
polarization is crucial.

This method has proved to be very successful in modelling the jet
flows in bright, nearby (FR\,I) radio galaxies, as we describe in
Section~\ref{weak}.  Given the derived velocity fields (and the
assumption of pressure equilibrium with the surroundings on large
scales), we can also use the laws of conservation of mass, momentum
and energy to estimate the energy and mass fluxes and the run of
entrainment rate along the jets \citep{LB02b}.  The magnetic-field
topology is not completely determined, however. The models show that
longitudinal and toroidal components dominate (at small and large
distances, respectively). The observed symmetry of the transverse
profiles of $I$ and $Q$ tells us that the field structure is not a
simple `grand design' helix on these scales \citep{LCB}, but leaves
open the question of whether the toroidal component is
vector-ordered. If it is, there should be systematic gradients in
Faraday rotation across the jets (we know that thermal electrons must
be entrained even if they are not present near the AGN).  Such
searches have so far been frustrated by large fluctuations in
foreground Faraday rotation.
  
The observational requirements for this technique are images of both
jets with good fidelity, at least 5 and preferably $>$10 beams across
their widths and noise levels low enough to detect 10\% polarization
at $4\sigma$ everywhere.  It must also be possible to correct the
effects of Faraday rotation (and depolarization) to high accuracy.

These methods have so far managed to provide answers to many of the
questions posed in Section~\ref{whybother}, but only for a small number of
  carefully selected sources. We have been restricted to jets which
  are bright and easily resolved. Thus we sample a limited
  (intermediate) range of luminosity, and can only model jets with
  large opening angles. We know very little about more powerful,
  highly-collimated jets, intrinsically weak sources or the systematic
  dependences of jet physics on power and environment.  To do better,
  we need much large samples of low-luminosity sources (high
  sensitivity at moderate resolution) and an instrument capable of
  much higher resolution and very high sensitivity for the powerful
  sources. We explore these applications in the following two
  sections.

\section{Decelerating jets: SKA1-MID}
\label{weak}

\subsection{Current results}

\begin{figure}
    \centering
    \includegraphics[width=0.85\textwidth]{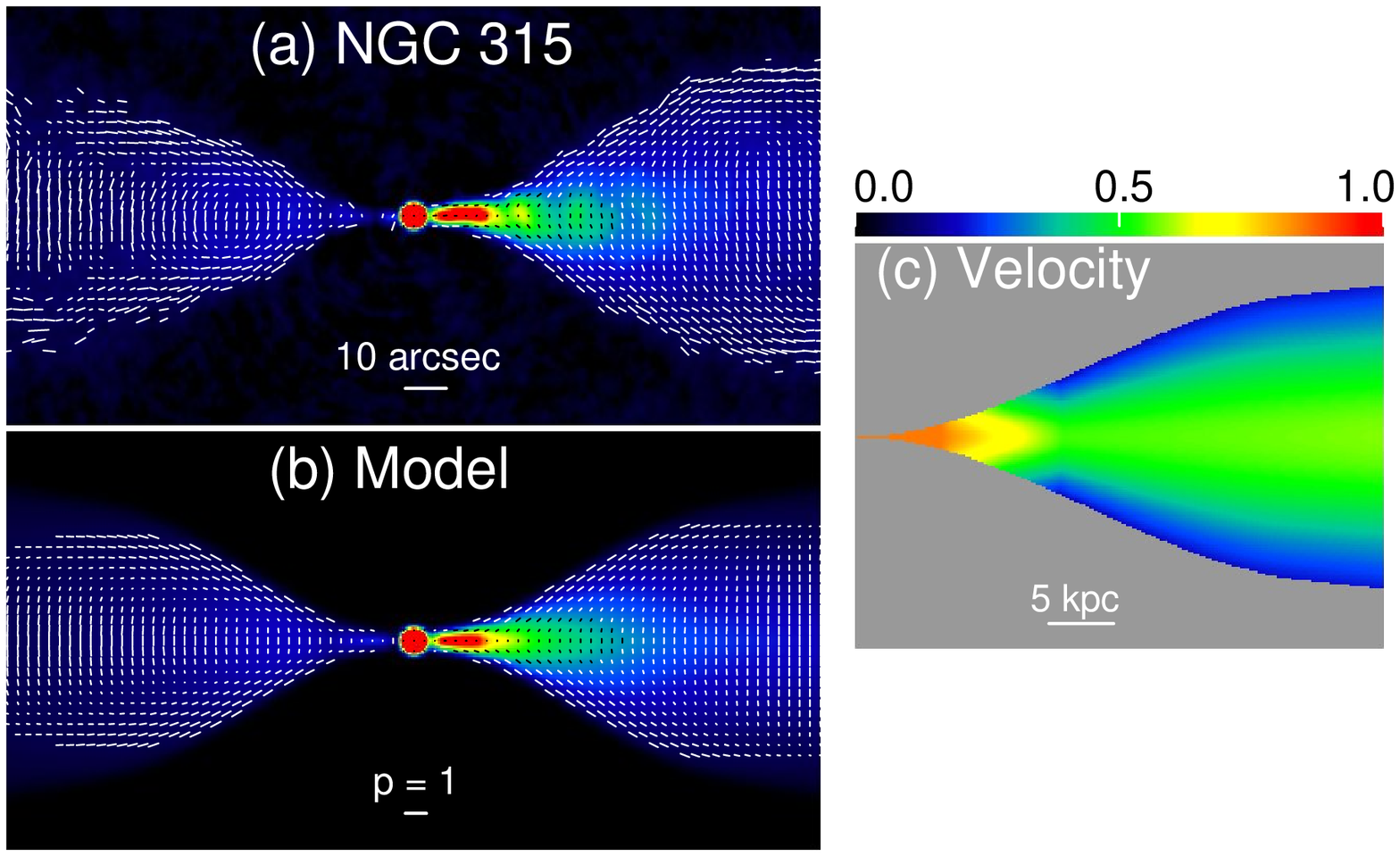}
    \caption{Observations and models of the nearby radio galaxy
      NGC\,315 \citep{LB14}. (a) Vectors with magnitudes proportional
      to the degree of polarization, $p$, and directions along the
      apparent magnetic field superposed on a false-colour plot of
      total intensity. (b) Model fit corresponding to (a). (c) The
      derived velocity field, in units of $c$.}
    \label{fig:model_example}
\end{figure}

Our current results on decelerating, FR\,I jets are summarised in
\cite{LB14}, and an example is shown in Fig.~\ref{fig:model_example}.
All of the jets first increase in opening angle and then recollimate to form conical flows at a
fiducial distance $r_0$ (the {\em flaring region}) in the range 2 -- 35\,kpc from the AGN.  At
$\approx$0.1$r_0$, the jets brighten abruptly at the onset of a
high-emissivity region and we find an outflow speed of
$\approx$0.8$c$, with a uniform transverse profile. Jet deceleration
first becomes detectable at $\approx$0.2$r_0$ and the outflow often
becomes slower at its edges than it is on-axis. Deceleration continues
until $\approx$0.6$r_0$, after which the outflow speed is usually
constant. The dominant magnetic-field component is longitudinal close
to the AGN and toroidal after recollimation, but the field evolution
is initially much slower than predicted by flux-freezing.  In the
flaring region, acceleration of ultrarelativistic particles is
required to counterbalance the effects of adiabatic losses and account
for observed X-ray synchrotron emission, but the brightness evolution
of the outer jets is consistent with adiabatic losses alone. These
results are best interpreted as effects of the interaction between the
jets and their surroundings.  The initial increase in brightness
occurs in a rapidly falling external pressure gradient in a hot,
dense, kpc-scale corona around the AGN.  We interpret the
high-emissivity region as the base of a transonic `spine' and suggest
that a subsonic shear layer starts to penetrate the flow there.  Most
of the resulting entrainment must occur before the jets start to
recollimate.

\subsection{SKA1-MID observations}

While the basic process of jet deceleration in twin-jet sources has
become clearer, there are still many unanswered questions. Almost all
of the sources we have modelled so far have luminosities close to $P
\approx 10^{24.5}$\,WHz$^{-1}$ at 1.4\,GHz.  The next set of questions
includes the following.
\begin{enumerate}
\item How does the characteristic recollimation scale $r_0$ vary with
  luminosity (jet power) and external pressure? Can we develop a way
  of predicting a priori which jets will decelerate and the form of
  their outer structures?
\item What are the distributions of energy and mass flux? How do these
  depend on black hole mass, accretion rate and ISM/IGM properties?
  Do they correlate with estimates from the size and pressure of
  jet-blown `cavities' in the IGM/ICM?  Can we infer
  the energy and momentum input integrated over the source population
  for comparison with galaxy formation models?
\item What causes jet emission to `turn on' at the flaring point,
  $\approx 0.1r_0$?
\item How important is stellar mass input in decelerating the weakest
  jets (it can be differentiated from boundary-layer entrainment
  because no slow boundary layer is formed)?
\item Can we decide unequivocally whether the toroidal field component
  is vector ordered and, if so, does it contribute to jet collimation
  on kpc scales? [To do this, we will need to measure Faraday rotation
    for a sample of sources selected to have low foreground contamination.]
\item How important are intrinsic and environmental asymmetries? This
  cannot be assessed for an individual object, but relativistic and
  environmental asymmetries are expected to be uncorrelated, so we can
  separate their effects statistically for a sufficiently large
  sample.
\end{enumerate}
In order to answer these questions with any statistical significance,
we will need to increase the number of modelled sources from 10 to at
least a few hundred and to identify sub-samples with different
luminosities and environments.  Our long VLA observations typically
reach rms noise levels between 5 and 10$\mu$Jy/beam at angular
resolutions between 0.25 and 2.5\,arcsec (The upgraded Karl G. Jansky VLA is allowing us to go up
to a factor of 5 deeper, but only with long multi-configuration
observations in special cases).  For an integral source count $N(>S)
\propto S^{-1.5}$ (appropriate for these nearby sources), we would
need to increase the sensitivity by a factor of 10 from the flux
densities alone (i.e.\ an rms noise level of $0.5 - 1.0\mu$Jy/beam).
Detailed trade-offs depend on source brightness, size and
environment. For example: the need to measure low Faraday rotations
argues for lower frequencies, but we cannot tolerate a significant
variation of rotation across the observing beam; there are systematic
variations of jet size (and therefore required resolution) with
luminosity, and so on.  Band 3 is therefore preferred for the
measurement of low Faraday depths; Band 5 for jet modelling. These
observations would also probe the magnetoionic environments of the
radio galaxies in great detail \citep{3c31RM,bands}.

An initial estimate is that we would need $\approx$300 observations
(mostly pointed, but perhaps also obtained from surveys) with high
image fidelity, resolution $\approx$0.25 \,arcsec and rms noise level
$\la$1\,$\mu$Jy/beam in SKA1-MID, Band 5.  For the combined SKA1-MID
array \citep[Table 9]{baseline} and allowing for a factor of 2
increase in noise from weighting \citep{imaging}, this would give an
average of $\ga$50\,min on-source time per observation, or $\ga$240\,hr in total.
The observations would still be feasible in an Early Science phase,
but longer integrations (by factors $\ga$4) or correspondingly smaller
sample sizes would be required.

\section{Fast, powerful jets: SKA2}
\label{powerful}

\begin{figure}
\includegraphics[width=.95\textwidth]{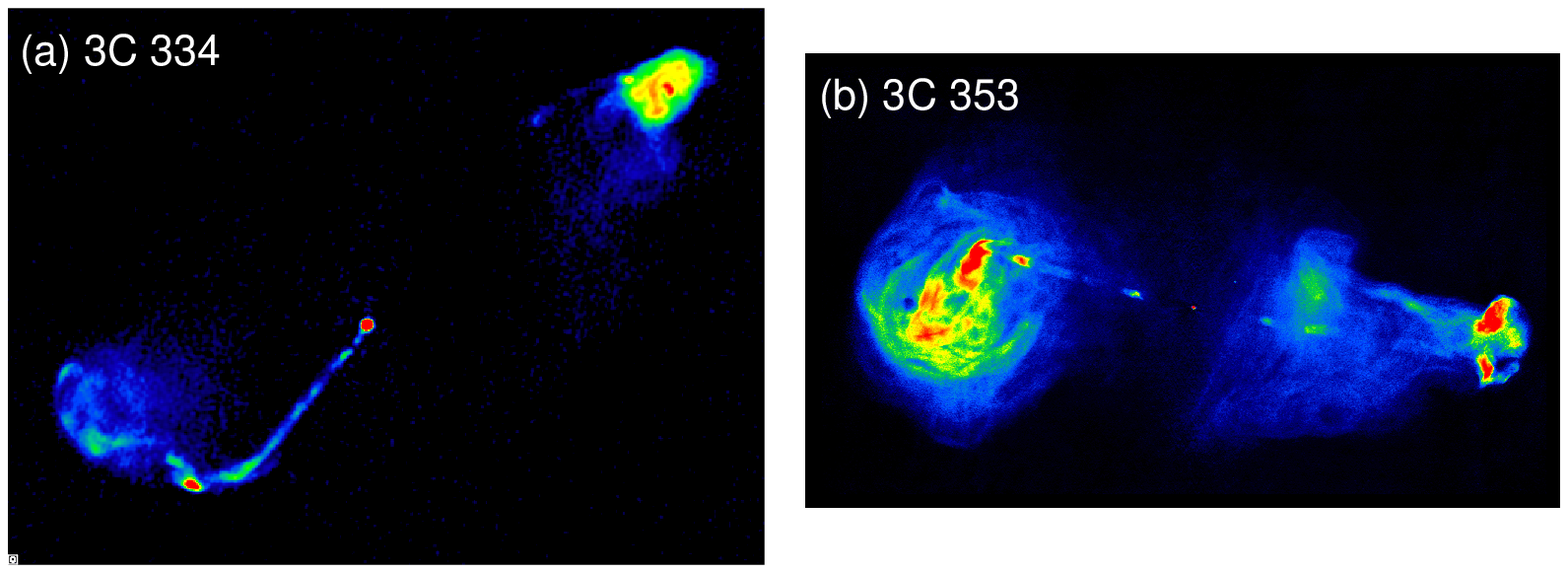}
\caption{VLA images of (a) the powerful quasar 3C\,334 \citep{Bridle94}
  and (b) the radio galaxy 3C\,353 \citep{SBB}.  Note the
  very narrow strong-flavour jets. 3C\,334 is thought to be at $\la
  40^\circ$ to the line of sight and shows a pronounced
  jet/counter-jet asymmetry (the counter-jet is only visible at large
  distances from the AGN). In contrast, 3C\,353 is close to the
  plane of the sky and both jets are visible, but its asymmetries are
  dominated by the environment.}
\label{fig:3c334_353}
\end{figure}

The strong-flavour jets found in powerful radio galaxies and quasars,
particularly those associated with FR\,II sources, are far harder to
study, as they are narrow and (in the case of counter-jets) also very
faint (e.g. Fig.~\ref{fig:3c334_353}).

We aim to distinguish between two hypotheses: (a) strong jets are only
mildly relativistic ($\beta \approx 0.5 - 0.7$; $\Gamma \approx 1.2 -
1.4$) as inferred naively from the statistics of their integrated
sidedness ratios \citep{WA,Mullin} or (b) they have a two-component
velocity structure, with a much faster ($\beta > 0.95$) spine
surrounded by a slower shear layer that dominates the emission except
in very end-on cases.  Case (b) has long been suspected \citep{Bridle94}, but never
proved: arguments in favour include the following.
\begin{enumerate}
\item High pattern speeds are inferred from VLBI proper-motion
  measurements for pc-scale jets in flat-spectrum quasars (the end-on
  counterparts of sources like 3C\,334;  
  \citealt{MOJAVE}). If these are representative of the bulk motion, then
  flow Lorentz factors up to $\Gamma \approx 40$ are
  required. Deceleration of such powerful jets to mildly relativistic
  speeds on kpc scales cannot occur without disruption.
\item If the X-rays observed in the extended jets associated with
  core-dominated sources are generated by inverse Compton scattering
  of cosmic microwave background photons by relativistic electrons in
  the jet, then the bulk Lorentz factors must be large 
  ($\Gamma \ga 10$) and the jets must be close to the line of sight
  \citep{Tavecchio2000,Celotti2001}. Note, however, that there are
  arguments against this interpretation of the X-ray
  emission, at least in some sources \citep{Hard06,MG14}.
\item The jet/counter-jet ratio and observations of apparent
  superluminal speeds in the jet of M\,87 both indicate $\Gamma
  \ga 6$ over at least part of its length \citep{m87ref}.
\end{enumerate}

\begin{figure}
\includegraphics[scale=0.8]{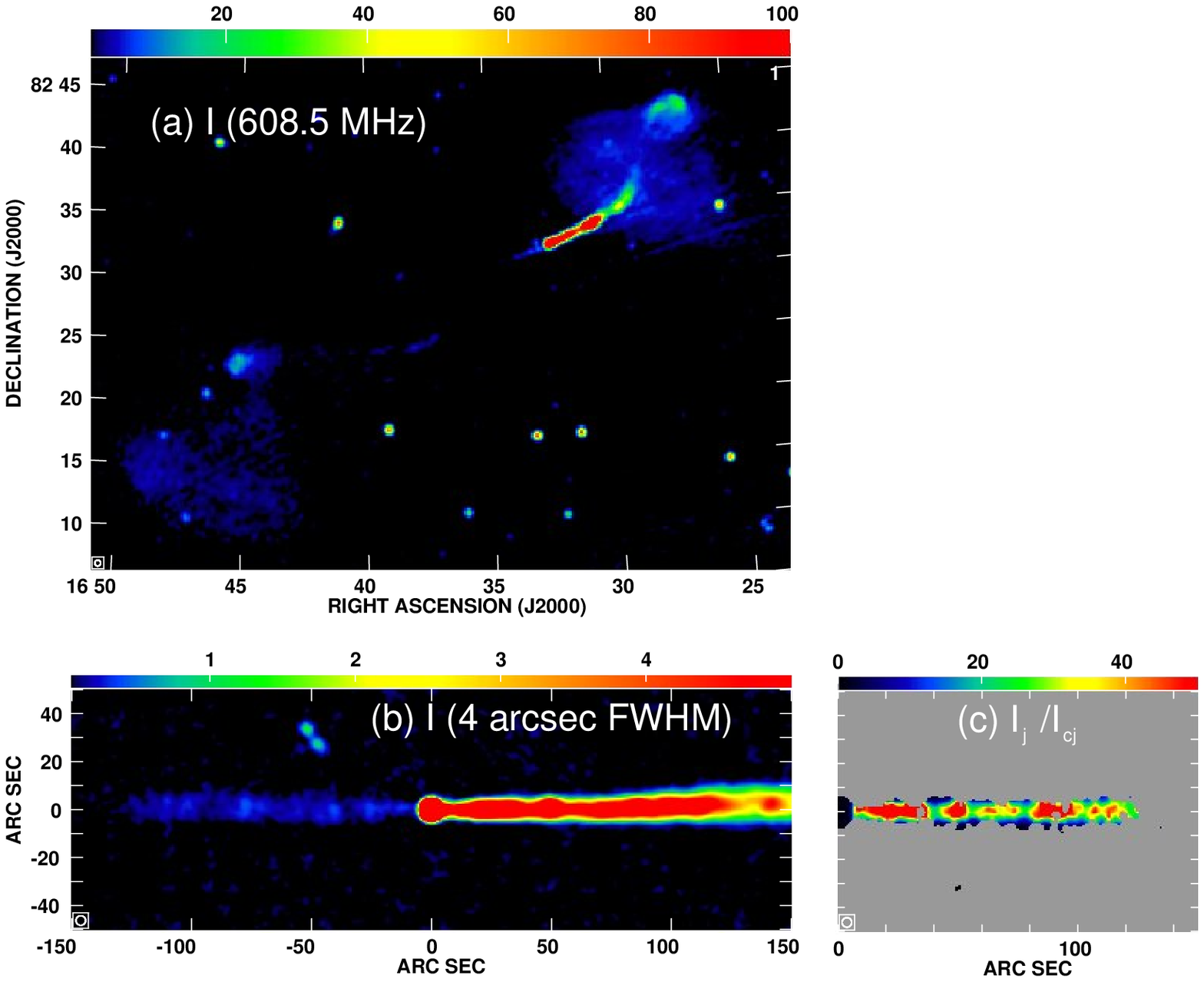}
\caption{Images of NGC\,6251. (a) Overall structure (WSRT, 609\,MHz, from \citealt{Mack97}).
  (b) Rotated image of the jets at 1665\,MHz, 4\,arcsec FWHM from archival VLA data. (c)
  Sidedness ratio $R = I_{\rm j}/I_{\rm cj}$, derived from (b) and 
  blanked (grey) where $I < 3\sigma$ in either jet.  The jets in this
  source represent a transition between weak and strong flavours: they
  flare, but still maintain a large side-to-side ratio. The sidedness
  ratio image in panel (c) strongly suggests a substantial velocity
  gradient across the jets. \label{fig:ngc6251}}
\end{figure}

There are a number of transition cases between weak and strong-flavour
jets which do not appear to decelerate significantly but which have
counter-jets that can now be imaged with the VLA. The best example is
NGC\,6251 (Fig~\ref{fig:ngc6251}), for which observations are under way.

Modelling of true strong-flavour jets such as those found in luminous
quasars is a far more challenging prospect, even for SKA2.
Counter-jets can currently only be imaged with something approaching adequate
s/n in sources such as 3C\,353 (Fig.~\ref{fig:3c334_353}, \citealt{SBB}) which are very close to the
plane of the sky (in which case our modelling technique does not
work).  We really need to image both jets in powerful sources which
are reasonably inclined.  As an example, if the straight part of the
counter-jet in 3C\,334 (Fig.~\ref{fig:3c334_353}) has $I \approx 10
\mu$Jy/beam at 5\,GHz with a beam of 0.35\,arcsec (just below the
current limit) and a width of 0.2 -- 0.6\,arcsec, we would need a
resolution of $\approx$0.05\,arcsec, and an rms noise level of $\la
10$\,nJy to be able to image and model the counter-jet in linear
polarization.  In turn, this requires a dynamic range of $\ga 10^7:1$
(peak:rms).  Both requirements could be relaxed if adequate resolution
could be obtained at a lower frequency.

\section{Summary}

We have outlined some of the ways in which SKA could contribute to the
study of large-scale jets in radio galaxies. Extension of current work
on low and intermediate luminosity sources to large samples is
feasible for SKA1-MID. The case of powerful jets is an interesting
challenge for SKA2.

\bibliographystyle{apj}

\end{document}